\documentclass[twocolumn, osajnl, showpacs, superscriptaddress]{revtex4}  
\usepackage{graphicx}

\begin{document}
\title{Amplitude squeezed fiber Bragg grating solitons}
\author{Ray-Kuang Lee}
\affiliation{Institute of Electro-Optical Engineering, National Chiao-Tung University, Hsinchu, Taiwan}
\affiliation{National Center for High-Performance Computing, Hsinchu, Taiwan}
\author{Yinchieh Lai}
\affiliation{Institute of Electro-Optical Engineering, National Chiao-Tung University, Hsinchu, Taiwan}
\date{\today}
\begin{abstract}
Quantum fluctuations of optical fiber Bragg grating solitons are investigated numerically by the back-propagation method. It is found for the first time that the bandgap effects of the grating act as a nonlinear filter and cause the soliton to be amplitude squeezed. The squeezing ratio saturates after a certain grating length and the fundamental Bragg soliton produces the optimal squeezing ratio.
\end{abstract}
\maketitle

In the literature, various types of optical soliton phenomena have been studied extensively in the area of nonlinear optical physics.
These include the nonlinear Schr{\" o}dinger solitons in dispersive optical fibers, spatial and vortex solitons in photorefractive materials/waveguides, and cavity solitons in resonators \cite{STrillo01}.
It has also been well known that fiber Bragg gratings (FBGs) with Kerr nonlinearity can exhibit optical soliton-like phenomena known as the Bragg grating solitons \cite{AAceves89, BEggleton96}.
The FBGs are one-dimensional photonic bandgap crystals with weak index modulation.
By utilizing the high dispersion of the FBGs near the bandedges, one can produce optical solitons in the anomalous dispersion side if the input pulse have suitable pulse-width and peak intensity.
From the theoretical point of view, solitary waves in one-dimensional periodic structures can travel with different group velocities and have been verified in some experiments \cite{BEggleton96}.
Even for two- or three-dimensional nonlinear photonic bandgap crystals, solitary waves can also exist \cite{ASukhorukov01} and have been observed recently \cite{Fleischer03}.

Most of the previous studies on Bragg grating solitons have been on the classical effects and there is almost no result on their quantum properties.
The quantum theory of traveling-wave optical solitons has been intensively developed during the past 15 years and several approaches have been successfully carried out to calculate the quantum properties of different traveling-wave optical solitons including the family of nonlinear Schr{\" o}dinger solitons \cite{Drummond87, Lai89a} as well as the self-induced-transparency solitons \cite{Lai90}.
Bragg solitons belong to the class of bi-directional pulse propagation problems where the quantum theory is still lack of enough consideration.
It is the aim of this study to bridge this gap by developing a general quantum theory for bi-directional pulse propagation problems and particularly applying the theory to the case of Bragg grating solitons.
It will be shown that the output Bragg soliton pulses will quantum-mechanically get amplitude-squeezed and the squeezing ratio can be calculated theoretically.

In our modeling, we use the nonlinear coupled mode equations (NCMEs) to describe the two bi-directional waves propagating in a uniform FBG.
We use the linearization approach to study the quantum effects of optical solitons in FBGs by extending the back-propagation method we previously developed \cite{YLai95} to the cases of nonlinear bi-directional propagation problems.
By following the same spirit of the back-propagation method, we will first derive a set of linear adjoint equations from the linearized NCMEs in such a way that any inner product between the solutions of the two equation sets is conserved during the time evolution.
Under the linearization approximation, the measurements performed after the time evolution can also be expressed in terms of the inner product between the perturbed quantum field operator and a measurement characteristic function.
By back-propagating the measurement characteristic function to $t=0$ through the solution of the adjoint equations, we can express the measured operator in terms of the input field operators which have known quantum characteristics.
In this way, the variance of the measured operator as well as its squeezing ratio can be calculated readily for a given measurement characteristic function. 

To be more explicit, let us consider the wave propagation problem in a one dimension fiber grating structure with the nonlinearity coming from the third order nonlinearity of the optical fiber.
With the self-phase modulation and cross-phase modulation effects, we model Bragg solitons by using the following NCMEs that describe the coupling between the forward and the backward propagating waves in a uniform FBG.
\begin{eqnarray}
\label{eqNCME1}
\frac{1}{v_g}\frac{\partial}{\partial t} U_a(z,t)+\frac{\partial}{\partial z} U_a &=& \\\nonumber
&&\hspace{-1in} i\delta U_a+i\kappa U_b+i\Gamma|U_a|^2 U_a+2 i\Gamma|U_b|^2 U_a\\
\label{eqNCME2}
\frac{1}{v_g}\frac{\partial}{\partial t} U_b(z,t)-\frac{\partial}{\partial z} U_b &=& \\\nonumber
&&\hspace{-1in} i\delta U_b+i\kappa U_a+i\Gamma|U_b|^2 U_b+2 i\Gamma|U_a|^2 U_b
\end{eqnarray}
Here $U_{a,b}(z,t)$ represent the forward and backward propagation pulses respectively,
$v_g $ is the group velocity of the pulse,
$\kappa$ is the coupling coefficient, $\lambda_B$ is the Bragg wavelength, 
$\delta $ is the wavelength detuning parameter, and $\Gamma$ represents the self-phase modulation coefficient.
\begin{figure}
\begin{center}
\includegraphics[width=3.0in, height=2.0in]{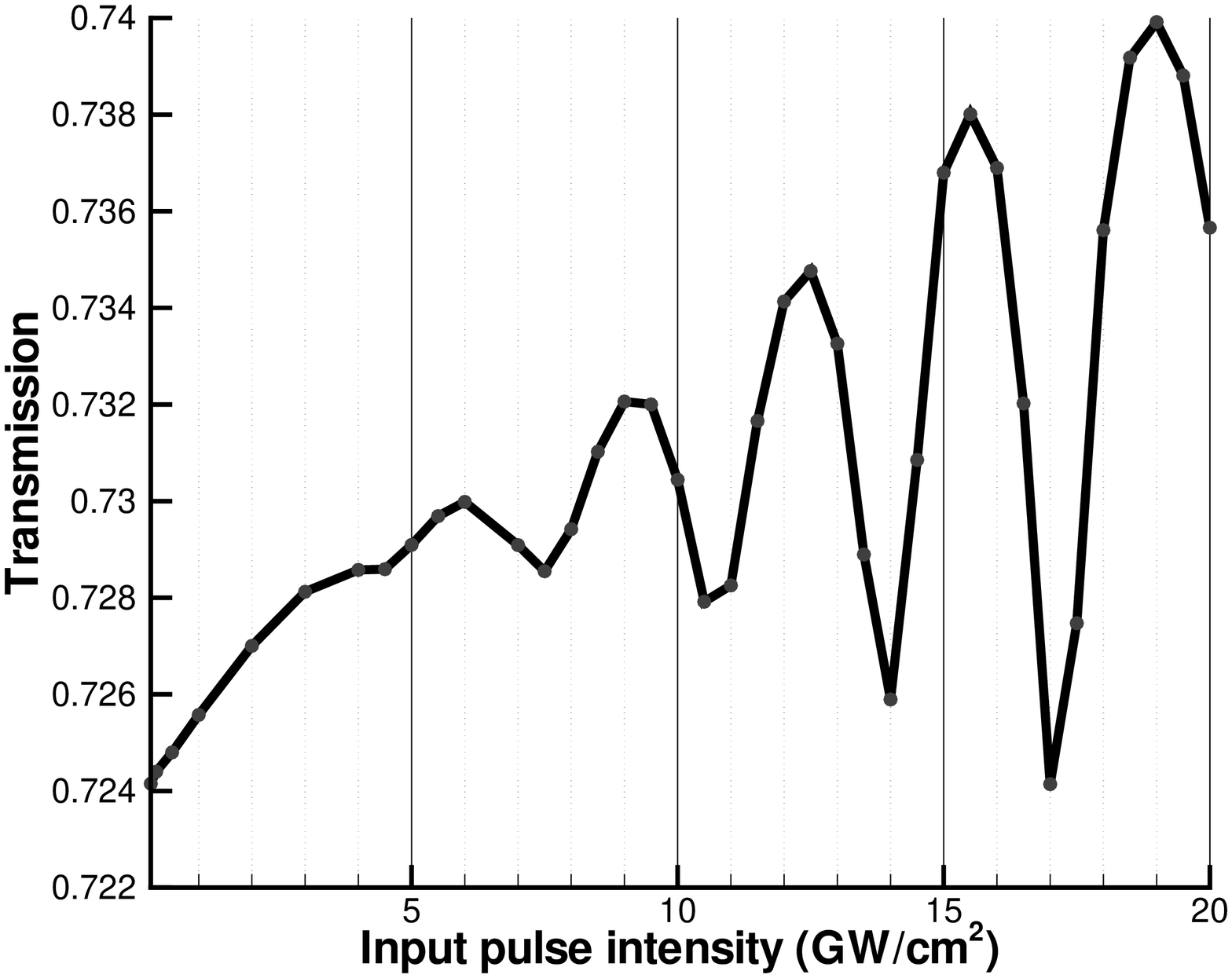} 
\includegraphics[width=3.0in, height=2.0in]{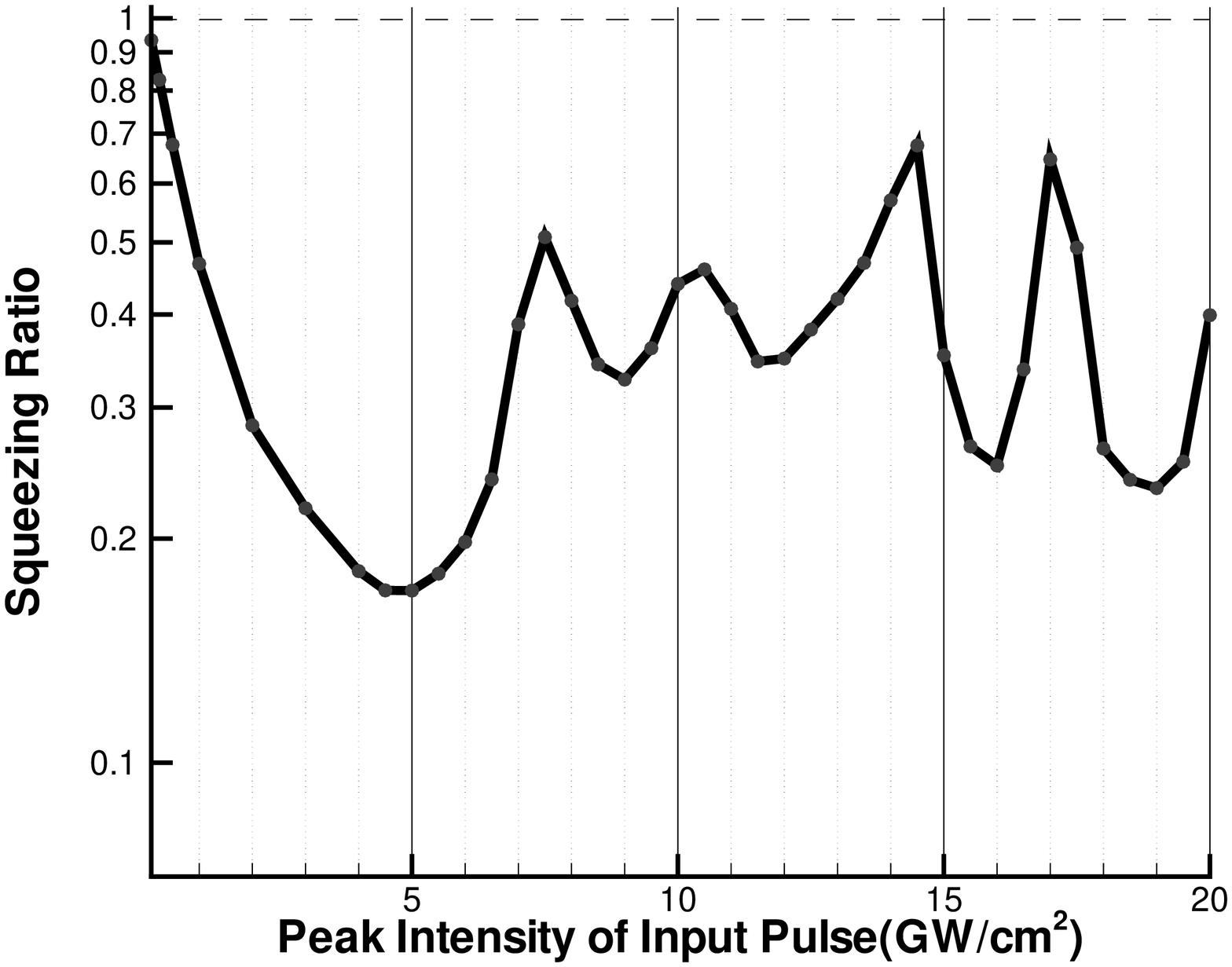} 
\caption{Transmission (top) and optimal squeezing ratio (bottom) for Bragg solitons with different input intensities.}
\label{SRIfig}
\end{center}
\end{figure}
This set of NCMEs has approximate analytical soliton solutions for the case of infinite grating length, as is shown by Aceves and Wabnitz with the introduction of the massive Thirring model \cite{AAceves89}. However, for gratings of finite length, no analytic solution can be found.
So in our studies we directly use the finite difference numerical simulation method with the parameters based on the first experimental report in the literature \cite{BEggleton96}. 
We consider a 60 $ps$ FWHM {\it sech}-shaped pulse incidents into a uniform grating with 15.0 $cm^{-1}$ wavenumber detuning from the center of the bandgap.
The coupling strength of the fiber grating is 10 $cm^{-1}$. 
When the input intensity is below the required value for forming a solitary pulse in the FBGs (about 4.5 $GW/cm^2$ in this case), the peak intensity of the pulse will decrease along the propagation.
On the other hand, when the input intensity is above 4.5 $GW/cm^2$, the peak intensity oscillates.
Only when the nonlinearity can exactly compensate the dispersion induced by the FBGs, we can have a stable solitary pulse inside the grating.
After obtaining these classical solutions, we now turn to the calculation of their quantum properties.

Since for optical solitons the average photon number is usually very large, we can safely use the linearization approximation to study their quantum effects.
By setting
\begin{eqnarray*}
U_a(z,t) &=& U_{a0}(z,t) +\hat{u}_a(z,t)\\
U_b(z,t) &=& U_{b0}(z,t) +\hat{u}_b(z,t)
\end{eqnarray*}
and substituting them into Eq. (\ref{eqNCME1}-\ref{eqNCME2}) for linearization, we obtain the linear quantum operator equations in Eq. (\ref{ptbeq}) that describe the evolution of the quantum fluctuations associated with the Bragg solitons.
The quantum perturbation fields $\hat{u}_a(z,t)$ and $\hat{u}_b(z,t)$ have to satisfy the following equal time commutation relations:
\begin{eqnarray*}
&&[\hat{u}_a(z_1,t),\hat{u}^\dag_a (z_2,t)] = [\hat{u}_b(z_1,t),\hat{u}^\dag_b (z_2,t)] = \delta (z_1-z_2)\\
&&{[}\hat{u}_{a,b}(z_1,t),\hat{u}_{a,b} (z_2,t)] = [\hat{u}^\dag_{a,b}(z_1,t),\hat{u}_{a,b}^\dag(z_2,t)] = 0
\end{eqnarray*}
\begin{widetext}
\begin{eqnarray}
\label{ptbeq}
\frac{1}{v_g}\frac{\partial}{\partial t} \left(\begin{array}{c}\hat{u}_a\\\hat{u}_b\end{array}\right) &=& \left(\begin{array}{cc} -\frac{\partial}{\partial z}+i\delta+2 i\Gamma|U_{a0}|^2+2 i\Gamma|U_{b0}|^2
& i\kappa+2 i\Gamma U_{a0}U_{b0}^\ast\\
i\kappa+2 i\Gamma U_{a0}^\ast U_{b0} & \frac{\partial}{\partial z}+i\delta+2 i\Gamma|U_{a0}|^2+2 i\Gamma|U_{b0}|^2\end{array}\right)\left(\begin{array}{c}\hat{u}_a \\ \hat{u}_b \end{array}\right)\\\nonumber
&+& \left(\begin{array}{cc} i\Gamma U_{a0}^2 & 2 i\Gamma U_{a0} U_{b0}\\
2i\Gamma U_{a0}U_{b0} & i\Gamma U_{b0}^2 \end{array}\right)\left(\begin{array}{c}\hat{u}^\dag_a \\ \hat{u}^\dag_b \end{array}\right)\\
\label{adjeq}
\frac{1}{v_g}\frac{\partial}{\partial t}\left(\begin{array}{c}u^A_a\\u^A_b\end{array}\right) &=&
\left(\begin{array}{cc} -\frac{\partial}{\partial z}+i\delta+2 i\Gamma|U_{a0}|^2+2 i\Gamma|U_{b0}|^2
& i\kappa+2 i\Gamma U_{a0} U_{b0}^\ast\\
i\kappa+2 i\Gamma U_{a0}^\ast U_{b0} & \frac{\partial}{\partial z}+i\delta+2 i\Gamma|U_{a0}|^2+2 i\Gamma|U_{b0}|^2\end{array}\right)\left(\begin{array}{c}u^A_a \\ u^A_b \end{array}\right)\nonumber\\
&+& \left(\begin{array}{cc} -i\Gamma U_{a0}^2 & -2 i\Gamma U_{a0} U_{b0}\\
-2i\Gamma U_{a0}U_{b0} & -i\Gamma U_{b0}^2 \end{array}\right)\left(\begin{array}{c} u^{A\ast}_a \\ u^{A\ast}_b \end{array}\right)
\end{eqnarray}
\end{widetext}
If we define the inner product operation according to 
\begin{eqnarray}
\langle\vec{f}|\vec{\hat{g}}\rangle &=& \frac{1}{2}\int dz\,[f^\ast_a \hat{g}_a+f_a \hat{g}^\dag_a+f^\ast_b \hat{g}_b+f_b \hat{g}^\dag_b]
\end{eqnarray}
and introduce the corresponding set of adjoint equations in Eq. (\ref{adjeq}), then it is easy to show that the inner product between the solutions of the two equation sets is preserved along the time axis, $\frac{d}{d t}\langle \vec{u}^A|\vec{\hat{u}}\rangle = 0$.
By taking advantage of this property, we can express the inner product of the output quantum perturbation operator with a projection function in terms of the input quantum operator.
This will allow us to calculate the quantum uncertainty for the inner product of the output quantum operator with any given projection function.
For the photon number measurement, the pulse shape of the projection function is simply the classical output solution \cite{HHaus90}and the squeezing ratio can be calculated according to:
\begin{eqnarray}
R(T) &=& \frac{var[\langle \vec{f}(z)| \vec{\hat{u}}(z,t=T)]}{var[\langle \vec{f}(z)| \vec{\hat{u}}(z,t=0)]}\\\nonumber
&=& \frac{var[\langle \vec{F}_T(z)| \vec{\hat{u}}(z,t=0)]}{var[\langle \vec{f}(z)| \vec{\hat{u}}(z,t=0)]}
\end{eqnarray}
Here $\vec{f}(z)$ is the original  projection function and $\vec{F}_T(z)$ is the back-propagated projection function.

In the following we shall present our results of simulation. The transmission of the FBGs with different input intensities for a constant FBG length (50 $cm$) is shown in the top curve of Fig. \ref{SRIfig}.
The calculated photon number squeezing ratio is shown in the bottom for the same parameters.
The squeezing ratio decreases monotonically when the intensity is smaller than the intensity of the fundamental soliton but will begin to oscillate strongly when the intensity is much larger than that of the fundamental soliton.
The oscillation behaviors of the FBG transmission and the squeezing ratio match very well. That is, the squeezing ratio has a local minimum when the transmission has a local maximum.
Intuitively the periodic grating structure acts like a spectral filter which can filter out the noisier high frequency components in the soliton spectrum and produce a net amplitude squeezing effect just as in the previous solution amplitude squeezing experiments where a spectral filter is cascaded after a nonlinear fiber \cite{Friberg96, TOpatrny02}. 
It is also intuitively clear that larger amplitude squeezing should occur when the transmission curve is saturated.
\begin{figure}
\begin{center}
\includegraphics[width=3in]{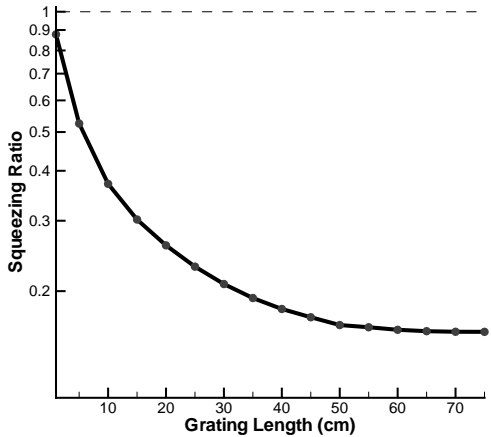} 
\caption{Squeezing ratio for Bragg solitons with different length of FBGs.}
\label{SRLfig}
\end{center}
\end{figure}
Fig. \ref{SRLfig} shows the dependence of the calculated squeezing ratios for different FBG lengths.
If we only consider the gratings with the length longer than 1 $cm$, we find that the squeezing ratio monotonically decreases with the FBG length and saturates at the length around 60 $cm$.
Intuitively this is because the filtering effect of the grating will unavoidably introduce additional noises on the light fields and eventually cause the squeezing ratio to become saturated.

To summarize, we have developed a general quantum theory for bi-directional nonlinear optical pulse propagation problems and have used it to calculate the squeezing ratio of Bragg solitons in one-dimensional photonic bandgap crystals for the first time.
We find that the output pulses can get amplitude squeezed and the squeezing ratio exhibits interesting relations with the fiber grating length as well as with the intensity of the input pulse. It will be very interesting to see if one can actually observe these effects experimentally in the future.


\begin{thebibliography}{}
\bibitem{STrillo01}
For a review of various soliton phenomena, please see
{\it ``Spatial Solitons''}, edited by S. Trillo and W. Torruellas (Springer, 2001).

\bibitem{AAceves89}
A. B. Aceves, and S. Wabnitz,
\pl A {\bf 141}, 37 (1989).

\bibitem{BEggleton96}
B. J. Eggleton, R. E. Slusher, C. M. de Sterke, P. A. Krug, and J. E. Sipe,
\prl {\bf 76}, 1627 (1996).

\bibitem{ASukhorukov01}
A. A. Sukhorukov and Y. S. Kivshar
\prl {\bf 87}, 083901 (2001).

\bibitem{Fleischer03}
J. W. Fleischer, M. Segev, N. K. Efremidis, and D. N. Christodoulides,
\nat {\bf 422}, 147 (2003).

\bibitem{Drummond87}
P. D. Drummond and S. J. Carter,
\josab {\bf 4}, 1565 (1987).

\bibitem{Lai89a}
Y. Lai and H. A. Haus,
\pra {\bf 40}, 844 (1989); \pra {\bf 40}, 854 (1989).

\bibitem{Lai90}
Y. Lai and H. A. Haus,
\pra {\bf 42}, 2925 (1990).

\bibitem{YLai95}
Y. Lai and S.-S. Yu,
\pra {\bf 51}, 817 (1995).

\bibitem{HHaus90}
H. A. Haus and Y. Lai
\josab {\bf 7}, 386 (1990).

\bibitem{Friberg96}
S. R. Friberg, S. Machida, M. J. Werner, A. Levanon, and T. Mukai,
\prl {\bf 77}, 3775 (1996).

\bibitem{TOpatrny02}
T. Opatrn{\' y}, N. Korolkova, and G. Leuchs,
\pra {\bf 66}, 053813 (2002).

\end{thebibliography}
\end{document}